\documentclass{jetpl}
\twocolumn

\title{Phase diagram of superfluid $^3$He in ''nematically ordered'' aerogel}

\rtitle{Phase diagram of superfluid $^3$He in ''nematically
ordered'' aerogel}

\sodtitle{Phase diagram of superfluid $^3$He in ''nematically
ordered'' aerogel}

\author{R.\,Sh.\,Askhadullin$^+$, V.\,V.\,Dmitriev\thanks{e-mail: dmitriev@kapitza.ras.ru}, D.\,A.\,Krasnikhin,
P.\,N.\,Martynov$^+$, A.\,A.\,Osipov$^+$, A.\,A.\,Senin,
A.\,N.\,Yudin}

\rauthor{R.\,Sh.\,Askhadullin$^+$, V.\,V.\,Dmitriev,
D.\,A.\,Krasnikhin, P.\,N.\,Martynov$^+$, A.\,A.\,Osipov$^+$,
A.\,A.\,Senin, A.\,N.\,Yudin}

\sodauthor{R.\,Sh.\,Askhadullin$^+$, V.\,V.\,Dmitriev,
D.\,A.\,Krasnikhin, P.\,N.\,Martynov$^+$, A.\,A.\,Osipov$^+$,
A.\,A.\,Senin, A.\,N.\,Yudin}

\address{P.\,L. Kapitza Institute for Physical Problems RAS,
2 Kosygina str., 119334 Moscow, Russia\\
$^+$ A.\,I.\,Leypunsky
Institute for Physics and Power Engineering, Obninsk, Kaluga
region, Russia}

\dates{11 February 2012}{*}

\abstract{

Results of experiments with liquid $^3$He immersed in a new type
of aerogel are described. This aerogel consists of
Al$_2$O$_3\cdot$H$_2$O strands which are nearly parallel to each
other, so we call it as a ``nematically ordered'' aerogel. At all
used pressures a superfluid transition was observed and a
superfluid phase diagram was measured. Possible structures of the
observed superfluid phases are discussed.}


\begin{document}
\maketitle
\section{Introduction}
An asymmetry of a volume filled by superfluid $^3$He can influence
on resulting pairing states. For example, in the case of a
restricted geometry, boundaries of the container can suppress some
components of the superfluid order parameter \cite{AGR}. This
distortion persists over a distance of the order of the
temperature dependent superfluid correlation length $\xi=\xi(T)$,
which diverges at the superfluid transition temperature. Theory
predicts, that restricted geometry may stabilize superfluid phases
which do not occur in bulk liquid $^3$He \cite{Barton, LiHo}. In
superfluid $^3$He inside a narrow gap (or in $^3$He film) a planar
type distortion is expected for the B phase in agreement with
results of recent experiments \cite{lev1,lev2}. A spatially
inhomogeneous order parameter with polar core may be realized in a
narrow channel. This prediction has not been unambiguously
confirmed by experiments, however measurements of mass
supercurrent in narrow channels indicate a possible phase
transition at the temperature just below the superfluid transition
temperature \cite{parp,pek}. It is probable that these
observations are associated with the transition into such kind of
polar-type superfluid phase.

Another way to introduce the anisotropy into superfluid $^3$He is
to use $^3$He confined in a globally anisotropic aerogel. It is
known that the high porosity silica aerogel does not completely
suppress the superfluidity of $^3$He \cite{Porto,Halp}. It is also
established that superfluid phases of $^3$He in aerogel (A-like
and B-like phases) are similar to superfluid phases of bulk $^3$He
(A and B phases respectively) if the anisotropy of the aerogel is
weak or if it corresponds to the squeezing deformation
\cite{Barker,Dmit1,Kun,Dmit2010,Halp2011}. In this case the
anisotropy of the aerogel influences only on the orientation of
the $^3$He superfluid order parameter and on its spatial structure
\cite{Kun,Dmit2010,Vol}. However, recent theoretical
investigations \cite{Aoyama} show that the stretching anisotropy
of the aerogel should result in a polar distortion of the A-like
phase of superfluid $^3$He in aerogel. Moreover, if the anisotropy
is large enough then, in some range of temperatures just below the
superfluid transition temperature, the pure polar phase may be
more favorable than the A phase. Unfortunately silica aerogels are
rather fragile, therefore in practice the stretching anisotropy
can be obtained only in process of aerogel preparation \cite{Pol}
and the large value of this anisotropy is hardly achievable.

In this paper we present results of nuclear magnetic resonance
(NMR) studies of liquid $^3$He confined in a new type of aerogel
\cite{Askh}. This aerogel consists of Al$_2$O$_3\cdot$H$_2$O
strands with a characteristic diameter $\sim$50\,nm and a
characteristic separation of $\sim$200\,nm (see Fig.1 and the SEM
photo in \cite{diff}). The remarkable feature of this aerogel is
that its strands are oriented along nearly the same direction at a
macroscopic distance ($\sim$3-5\,mm), i.e. this aerogel may be
considered as almost infinitely stretched. To emphasize this
property we call it as ``nematically ordered'' aerogel.

\begin{figure}[h]
\begin{center}
\includegraphics[scale=0.4]{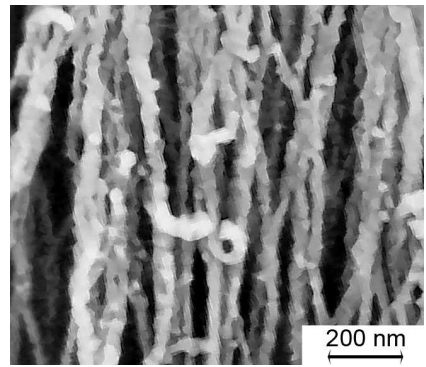}
\end{center}
\caption{Fig.1. The SEM photo of ``nematically ordered'' aerogel.}
\label{f1}
\end{figure}
\section{Experimental setup}
Our experimental chamber was made of epoxy resin "Stycast-1266"
and had two cylindric cells (see Fig.2). Two aerogel samples with
different porosities were placed freely in the cells. Samples had
a form of a cylinder with the diameter $\sim$4\,mm and with the
heights 2.6\,mm (the denser sample) and 3.2\,mm (the less dense
sample). Axes of the cylinders were oriented along aerogel strands
($\hat{\bf z}$-axis). Each cell was surrounded by transverse NMR
coil (not shown in Fig.2) with the axis along $\hat{\bf x}$.
Experiments were carried out in magnetic fields $\bf H$ from
106\,Oe up to 346\,Oe (the range of NMR frequencies was from
344\,kHz up to 1.12\,MHz) and at pressures from s.v.p. up to
29.3\,bar. We were able to rotate $\bf H$ by any angle $\mu$ in
the $\hat{\bf y}-\hat{\bf z}$ plane. Additional gradient coils
were used to compensate an inhomogeneity of $\bf H$ and to apply
the controlled field gradient. Residual inhomogeneity of $\bf H$
was $\sim 4\cdot10^{-5}$ for $\mu=0$ and $\sim 4\cdot10^{-4}$ for
$\mu=90^\circ$. About 30\% of the cell volumes were filled with
the bulk liquid, but usually it was easy to distinguish the signal
of superfluid $^3$He in aerogel from bulk $^3$He signal. Necessary
temperatures were obtained by a nuclear demagnetization cryostat
and were determined either by NMR in the A phase of bulk $^3$He
(when it was possible) or using a quartz tuning fork calibrated by
measurements of the Leggett frequency in bulk $^3$He-B. To avoid
paramagnetic signal from solid $^3$He aerogel samples have been
preplated by $\sim$2.5 atomic monolayers of $^4$He. The aerogel
strands have bends and their surface is rough so we assume that,
in spite of the preplating, a scattering of $^3$He quasiparticles
on the strands is diffusive.
\begin{figure}[t]
\begin{center}
\includegraphics[scale=0.25]{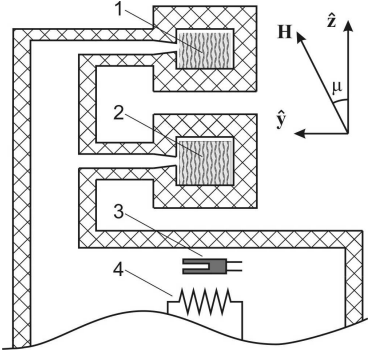}
\end{center}
\caption{Fig.2. The sketch of the experimental chamber. 1 -- the
denser sample; 2 -- the less dense sample; 3 -- quartz tuning
fork; 4 -- heater.} \label{f2}
\end{figure}

The described setup has been used also for measurements of spin
diffusion in normal liquid $^3$He in the same aerogel samples
\cite{diff}. The spin diffusion was found to be anisotropic in the
limit of low temperatures. Quasiparticles effective mean free
paths determined by aerogel strands were found to be:
$\lambda_{\parallel}\approx850$\,nm,
$\lambda_{\perp}\approx450$\,nm (for the denser sample) and
$\lambda_{\parallel}\approx1600$\,nm,
$\lambda_{\perp}\approx1100$\,nm (for the less dense sample). Here
$\lambda_{\parallel}$ and $\lambda_{\perp}$ are the mean free
paths along and normal to the aerogel strands respectively.

Most of the experiments described below were done with the denser
sample and the presented results were obtained using this sample
if not specially mentioned.

\section{Phase diagram}
On cooling from the normal phase we observed a superfluid
transition of $^3$He in both aerogel samples. The transition was
detected by continuous wave (CW) NMR with {\bf H} parallel to the
aerogel strands ($\mu=0$): at the transition temperature
($T_{ca}$) a positive NMR frequency shift ($\Delta \omega$) from
the Larmor value appears. The transition temperature is suppressed
in comparison with the superfluid transition temperature in bulk
$^3$He ($T_c$) and the suppression was found to be $\sim$2 times
less in the less dense sample than in the denser sample. The
pressure dependence of the suppression in terms of the superfluid
coherence length $\xi_0$ is shown in Fig.3.
\begin{figure}[b]
\begin{center}
\includegraphics[scale=0.8]{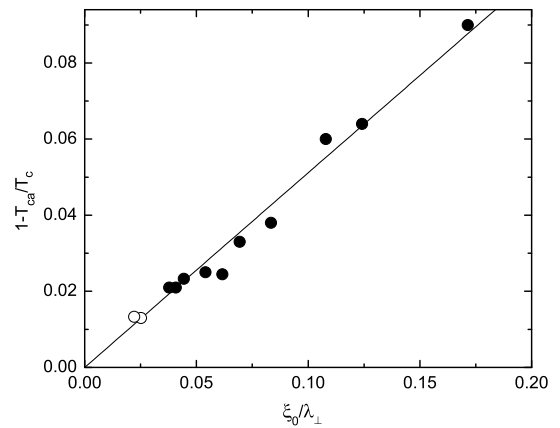}
\end{center}
\caption{Fig.3. The suppression of $T_{ca}$ of $^3$He in
``nematically ordered'' aerogel versus $\xi_0/\lambda_\perp$.
({\Large{$\circ$}}) -- the less dense sample, ({\Large$\bullet$})
-- the denser sample. The line is the best fit by $y=Ax$ with
$A=0.51$.} \label{f3}
\end{figure}
\begin{figure}[h]
\begin{center}
\includegraphics[scale=0.9]{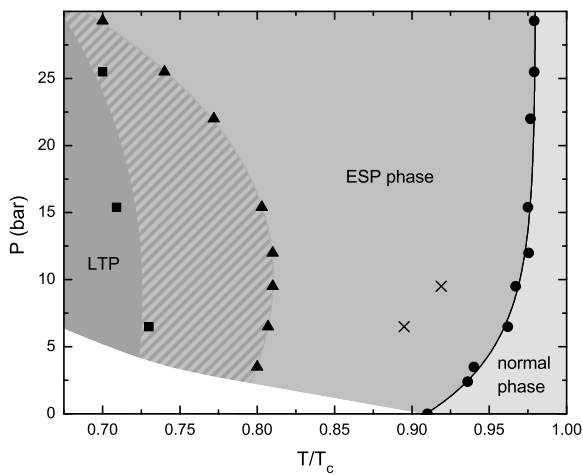}
\end{center}
\caption{Fig.4. The phase diagram of liquid $^3$He in
``nematically ordered'' aerogel obtained on cooling from the
normal phase. Note that the temperature is normalized to the
superfluid transition temperature in bulk $^3$He. See text for
explanations.} \label{f4}
\end{figure}

Fig.4 shows the measured phase diagram of superfluid $^3$He in
``nematically ordered'' aerogel. The filled circles correspond to
the transition from the normal phase into a ``high temperature''
superfluid phase. This phase belongs to a family of Equal Spin
Pairing (ESP) phases because its spin susceptibility is the same
as in the normal phase and does not depend on $T$. Below we call
this phase as the ESP1 phase. The triangles correspond to a
beginning of the 1st-order phase transition from the ESP1 phase
into a ``low temperature'' superfluid phase (LTP), where the
susceptibility is less than in the normal phase. A region of
coexistence of the LTP and the ESP1 phase is marked by a two-tone
area fill. Such a coexistence may be due to a pinning of the
interphase boundary on local inhomogeneities of the aerogel. The
squares correspond to the end of the transition into the LTP. On
subsequent warming the reverse 1st-order transition (from the LTP
into the ESP phase) is clearly visible only at $P\geq 12$\,bar and
begins at higher temperatures {\bf ($\sim 0.85\,T/T_c$)}. More
detailed description of the transitions and the superfluid phases
is given below.

We have found that NMR properties at high pressures
($P\geq$12\,bar) and at low pressures ($P\leq$6.5\,bar) are
qualitatively different. The high pressure behavior is illustrated
by Fig.5 where we present the temperature dependence of the
effective NMR frequency shift (2$\omega\Delta\omega$, where
$\omega$ is the NMR frequency) in continuous wave (CW) NMR
experiments at $P=29.3\,$bar and with $H\parallel {\bf \hat z}$.
On cooling from the normal phase we observed the transition into
the ESP1 phase with positive $\Delta \omega$ (open circles in
Fig.5). At $T\sim 0.7\,T_c$ the 1st-order transition into the LTP
starts. The frequency shift in this phase is larger than in the
ESP1 phase and on further cooling in some temperature range we
observe two NMR lines (from the LTP and from the ESP1 phase) with
different values of $\Delta\omega$. After the complete transition,
we observe on warming only the line from the LTP (filled circles
in Fig.5) until the reverse transition starts ($\sim
0.84\,T/T_c$). Surprisingly, the obtained in a such way the ESP
phase (we call it as ESP2 phase) is different from the ESP1 phase:
it has larger $\Delta \omega$ up to $T_{ca}$ (filled triangles in
Fig.5) and larger the NMR linewidth (insert of Fig.5). Being
obtained, this ESP2 phase remains stable down to $T\sim 0.7\,T_c$
(open triangles in Fig.5).
\begin{figure}[h]
\begin{center}
\includegraphics[scale=0.8]{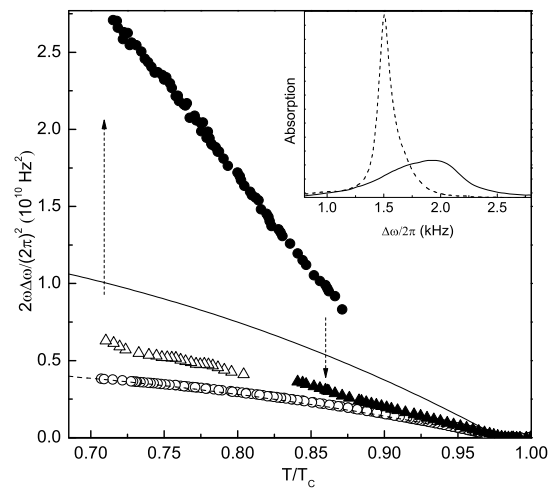}
\end{center}
\caption{Fig.5. The effective NMR frequency shift versus
temperature. $P=29.3$\,bar, $\mu=0$, $T_{ca}=0.979\,T_c$ and $H$=
346\,Oe. ({\Large{$\circ$}}) - the ESP1 phase; ({\Large$\bullet$})
- the LTP; ($\triangle$) - the ESP2 phase on cooling;
($\blacktriangle$) - the ESP2 phase on warming. Solid and dashed
lines - see Section 5 for the explanation. Insert: CW NMR
absorption lines in the ESP1 (dashed) and in the ESP2 phases
(solid) at the same temperature ($T=0.79\,T_c$). } \label{f5}
\end{figure}
\begin{figure}[h]
\begin{center}
\includegraphics[scale=0.8]{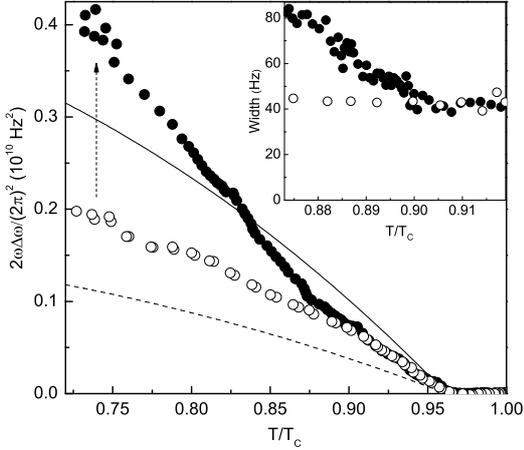}
\end{center}
\caption{Fig.6. The effective NMR frequency shift versus
temperature. $P=6.5$\,bar, $\mu=0$, $T_{ca}=0.962\,T_c$ and
$H$=346\,Oe. ({\Large{$\circ$}}) - the ESP1 phase;
({\Large$\bullet$}) - the LTP. Solid and dashed lines - see
Section 5 for the explanation. Insert: corresponding CW NMR
linewidths.} \label{f6}
\end{figure}

Fig.6 shows an example of the low pressure behavior. Here the
dependence of 2$\omega \Delta \omega$ on $T$ at $P=6.5$\,bar is
presented. On cooling from the normal phase we get the ESP1 phase
(open circles in Fig.6). Then, at $T\sim 0.8\,T_c$, the transition
into the LTP begins. This transition completes at $\sim 0.73\,T_c$
and on subsequent warming the LTP smoothly transforms into the ESP
phase (filled circles in Fig.6). This smooth transition ends at
$T_x=T\sim 0.91\,T_c$: above this temperature $\Delta\omega$ and
the linewidth are the same as in the ESP1 phase.

At intermediate pressures ($6.5\,$bar$<P<12\,$bar) we can not
distinguish between these two types of the behavior: no jump of
$\Delta \omega$ is seen, but the absolute value of the slope of
the dependence $\Delta \omega = \Delta \omega (T)$ essentially
increases just below $T_x$. At these pressures and near $T_x$ the
NMR frequency shifts in the LTP and ESP phases are close to each
other, so the final temperature width of the transition may mask a
jump of $\Delta \omega$. It is also possible that inhomogeneities
in the aerogel result in different types of the transition at
different places of the sample.

\section{Low temperature phase}
At high pressures the transition from the ESP phase into the LTP
is accompanied by a sharp decrease of the spin susceptibility and,
on further cooling, the susceptibility is decreasing. For ${\bf
H}\parallel \hat{\bf z}$ the NMR frequency shift in the LTP is a
few times larger than in the ESP phase and is close to the value
expected for bulk $^3$He-B with ${\bf \hat l}\perp{\bf H}$, where
${\bf \hat l}$ is an orbital vector oriented along the direction
of the gap anisotropy \cite{Vol2}. For example, at 29.3\,bar and
at $T=0.75\,T_{ca}$ the ratio of the observed shift and of the
expected bulk B phase shift is 0.86. Correspondingly, we assume
that at high pressures the order parameter of the LTP is close to
the parameter of bulk $^3$He-B and aerogel strands orient ${\bf
\hat l}$ normal to $\hat{\bf z}$.

At low pressures the LTP is not so close to bulk $^3$He-B.
Firstly, with decreasing pressure the above-mentioned ratio
decreases down to 0.4 at $P=$3.5\,bar. Secondly, the
susceptibility hardly changes during the transition from the ESP1
phase into the LTP.

The properties of the LTP may be explained if we suggest that the
order parameter corresponds to the order parameter of bulk
$^3$He-B with the polar distortion. Aerogel strands suppress the
gap in directions normal to their axes, i.e. normal to $\hat{\bf
z}$. If we choose the direction $\bf \hat l$ along $\bf \hat x$,
then the distorted B phase order parameter matrix averaged over
distances much larger than $\xi$ is:
\begin{equation}
\textbf{A}=\textbf{R}\begin{pmatrix}
b & 0 & 0 \\
0 & b & 0 \\
0 & 0 & a
\end{pmatrix}=\begin{pmatrix}
0 & 0 & aR_{13} \\
bR_{21} & bR_{22} & 0 \\
bR_{31} & bR_{32} & 0
\end{pmatrix},\label{eq1}
\end{equation}
where $\textbf{R}=\textbf{R}(\textbf{n},\Theta)$ is pure B phase
order parameter rotation matrix, $a$ and $b$ are positive and
${a^2+2b^2=1}$. If $a=b$ then we get pure Balian-Werthamer (BW)
state with isotropic energy gap (the case of bulk $^3$He-B in weak
magnetic field). Note that
$b/a={\Delta_{\perp}/\Delta_{\parallel}}<1$, where
$\Delta_{\perp}$ and $\Delta_{\parallel}$ are values of the gap in
directions normal and along $\bf \hat z$ respectively. The limit
$a=1$ and $b=0$ corresponds to the polar phase. The observed
temperature dependencies of $\Delta \omega$ show that $a$ and $b$
depend on temperature and pressure: the ratio $b/a$ is larger at
lower temperatures and at higher pressures.

\section{ESP phases}
Following \cite{Aoyama} we can suggest three variants for the ESP
phase: i) it can be analog of the A phase, i.e. its order
parameter corresponds to Anderson-Brinkman-Morel (ABM) model; ii)
it can be the A phase with the polar distortion; iii) it can be
the polar phase. The energy gap should be maximal along $\bf \hat
z$ and minimal in the $\bf{\hat{x}-\hat{y}}$ plane. If we choose
$\bf \hat x$ as a direction along which the gap is minimal then a
general form of the order parameter matrix for all three cases is:
\begin{equation}
\textbf{A}=\begin{pmatrix}
0 & ib & a \\
0 & 0 & 0 \\
0 & 0 & 0
\end{pmatrix},\label{eq2}
\end{equation}
where $a$ and $b$ are positive and ${a^2+b^2=1}$. Here we suggest
that the angle between $\bf \hat x$ and spin vector $\bf \hat d$
is zero. The pure ABM phase corresponds to $a=b$, while for the
pure polar phase $a=1$ and $b=0$.

We have carried out additional NMR experiments with ESP1 and ESP2
phases. In CW NMR experiments the dependence of $\Delta \omega$ on
$\mu$ was measured. In the denser sample we have found that
$\Delta \omega$ is positive for $\mu=0$ and equals zero for
$\mu=90^\circ$. In the less dense sample measurements have been
done also for $\mu=16^\circ$ and for $\mu=33^\circ$ and it was
found that $\Delta\omega\propto \cos^2 \mu$. In pulsed NMR
experiments the dependence of the initial frequency of a free
induction decay signal (FIDS) on the tipping angle of the
magnetization ($\beta$) was measured. It was found that
$\Delta\omega\propto \cos\beta$ for ${\bf H}\parallel {\bf \hat
z}$ and $\Delta\omega\propto (1-\cos\beta)$ for ${\bf H}\perp {\bf
\hat z}$ (see Fig.7). These results definitely exclude the
spatially homogeneous ABM (or distorted ABM) order parameter for
both ESP1 and ESP2 phases but qualitatively agree with an equation
for $\Delta \omega$ for the polar phase \cite{Aoyama}:
\begin{equation}
2\omega\Delta\omega =C\left(\cos\beta-\sin^2\mu \frac{5\cos\beta
-1}{4}\right), \label{eq3}
\end{equation}
where $C=\Omega_p^2$ and $\Omega_p$ is the Leggett frequency for
the polar phase. However, we should consider also another
possibility: our aerogel can be considered as almost infinitely
stretched and the ABM order parameter (as well as the distorted
ABM order parameter) can be in a two dimensional Larkin-Imry-Ma
(LIM) state. This state corresponds to a spatially inhomogeneous
distribution of $\bf \hat l$ in the $\bf {\hat x}-{\hat y}$ plane
\cite{Dmit2010,Vol}. A characteristic length of these
inhomogeneities should be less than the dipole length, because
well below $T_{ca}$ the observed CW NMR linewidths are small in
comparison with $\Delta \omega$. For the ABM phase in the LIM
state the dependence of $\Delta \omega$ on $\mu$ and $\beta$ is
also given by \eqref{eq3} (see Eq.15 in Ref.\cite{Dmit2010}) but
with different $C$. For a general case $C$ can be calculated in
weak coupling limit:
\begin{equation}
C =2\Omega_A^2\frac{3a^2-1}{3-4a^2b^2}, \label{eq4}
\end{equation}
where $\Omega_A$ is the Leggett frequency in the pure ABM phase.
For $a=b=1/\sqrt{2}$ (the ABM phase in the LIM state)
$C=\Omega_A^2/2$, while for $a=1,~b=0$ (the polar phase)
$C=4\Omega_A^2/3=\Omega_p^2$.
\begin{figure}[t]
\begin{center}
\includegraphics[scale=0.85]{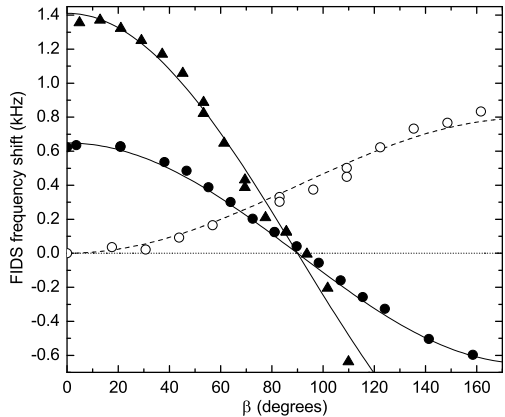}
\end{center}
\caption{Fig.7. The initial frequency of the FIDS versus $\beta$.
({\Large$\bullet$}):~the ESP1 phase, $\mu=0, P=9.5$\,bar,
$T=0.85\,T_c, ~H=346$\,Oe. ($\blacktriangle$): the ESP2 phase,
$\mu=0, P=29.3$\,bar, $T=0.835\,T_c, ~H=346$\,Oe.
({\Large{$\circ$}}): the ESP1 phase, $\mu=90^{\circ}, P=12$\,bar,
$T=0.87\,T_c, ~H=106$\,Oe. Solid lines and the dashed line are the
best fits by $\cos\beta$ and $(1-\cos \beta)$ dependencies
respectively.} \label{f7}
\end{figure}

Thus, our results indicate that ESP phases have the order
parameter \eqref{eq2} and, depending on $a$ and $b$, may
correspond either to the polar or to the LIM state of the ABM
phase with the polar distortion. ESP1 and ESP2 phases can exist at
the same conditions and presumably have different values of $a$
and $b$. The value of $\Delta \omega$ allows, in principal, to
find $C$ and to calculate $a$ and $b$: if $\mu=0$ and $\beta=0$,
then $C=2\omega\Delta\omega$. The problem is that $\Omega_A$ for
$^3$He in our aerogel is unknown. Therefore, in the first
approximation, we have used data for $\Omega_A$ in bulk $^3$He-A
\cite{Sch,Ah,Rand} rescaled in assumption that the relative
suppression of $\Omega_A$ equals the relative suppression of the
superfluid transition temperature ($T_{ca}/T_c$). Using these
values of $\Omega_A$ the expected dependencies of 2$\omega \Delta
\omega$ on temperature for the pure polar phase and for the pure
ABM phase in the LIM state have been calculated (dashed and solid
lines respectively in Figs.5 and 6). It is seen that at 29.3\,bar
the ESP1 phase is close to the pure ABM phase in the LIM state
while at 6.5\,bar the polar distortion is essential.

\section{Search for pure polar phase}
The polar distortion of the ABM order parameter depends on
temperature and is expected to be larger near $T_{ca}$
\cite{Aoyama}. To characterize this distortion we introduce a
parameter $K=C/\Omega_A^2$ and denote its value in the limit
$T\rightarrow T_{ca}$ as $K_0$. For the pure ABM phase in the LIM
state $K$ should be equal to 0.5 while for the pure polar phase we
expect $K=4/3$. The experimental dependence of $K_0$ on pressure
is shown in Fig.8. At low pressures $K_0 \approx 1.07$ (dashed
line in Fig.8) and from \eqref{eq4} we find that it corresponds to
a strongly distorted ABM phase with $a^2=0.73$ and $b^2=0.27$.
\begin{figure}[b]
\begin{center}
\includegraphics[scale=0.85]{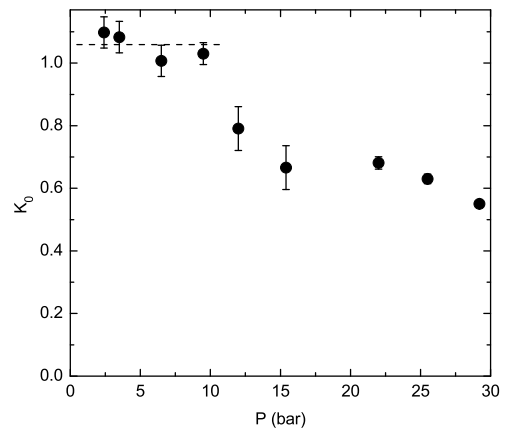}
\end{center}
\caption{Fig.8. The dependence of $K_0$ on pressure.} \label{f8}
\end{figure}
However, we can not exclude that the pure polar can be realized at
low pressures near $T_{ca}$. The point is that the suppression of
the bulk value of $\Omega_A$ may be larger than we have used in
our estimation of $C$ as it happens in the case of $^3$He in
standard silica aerogel \cite{Thun,Halp2008} or as it follows from
a slab model \cite{Thun2}. In fact at $P<$12\,bar an additional
suppression of $\Omega_A^2$ by 25\% is enough to get $K=4/3$. One
more argument in favor of the polar phase is that at low pressures
$K\geq 1$ in a finite range of temperatures below $T_{ca}$ and
starts to decrease on further cooling below some temperature
$T_k<T_{ca}$. The values of $T_k$ at 6.5\,bar and 9.5\,bar are
shown by crosses in Fig.4. At 2.4\,bar and 3.5\,bar the decrease
of $K$ have not been observed till the lowest obtained
temperatures. We note also that at low pressures the transition
from the LTP into the ESP1 phase is continuous. This is possible
if the distorted BW phase transforms into the polar phase (see
Eq.1 with $b\rightarrow 0$), while the transition into the ABM (or
into the distorted ABM) phase should be of the 1st-order as we
observe at high pressures.

\section{Conclusions}
We have measured the phase diagram of superfluid phases of $^3$He
in ``nematically ordered'' aerogel. Depending on conditions and
the prehistory 3 superfluid phases were observed: the LTP, the
ESP1 phase and the ESP2 phase. The LTP presumably has BW order
parameter with the polar distortion, while the ESP1 and the ESP2
phases correspond to the ABM order parameter with different values
of the polar distortion. There are indications that at low
pressures the pure polar phase may exist in some range of
temperatures just below $T_{ca}$. However, additional experiments
are necessary to check such a possibility.

Existing theoretical models for superfluid $^3$He in aerogel or
for the case of restricted geometry can not be directly applied to
our case. Inhomogeneous isotropic scattering model \cite{Thun}
which well describes $^3$He in nearly isotropic silica aerogel,
considers the isotropic scattering of quasiparticles and surely
can not be used to interpret our results. The theoretical model of
K.Aoyama and R.Ikeda \cite{Aoyama} of the A-like phase in
stretched aerogel is much closer to our situation and our results
for ESP phases qualitatively do not contradict this model.
However, the case of a strong anisotropy of the quasiparticles
mean free path has not been considered in \cite{Aoyama}. We also
note that the B phase with polar-type distortion has been
theoretically considered in detail only for the case of $^3$He
inside narrow cylindrical channels. Correspondingly, we think that
further theoretical investigations are necessary to explain the
observed properties of superfluid $^3$He in ``nematically
ordered'' aerogel.

We thank I.A.\,Fomin, W.P.\,Halperin, E.V.\,Surovtsev and
G.E.\,Volovik for useful discussions. This work was supported in
part by the Russian Foundation for Basic Research (grant \#
11-02-12069 ofi-m) and by the Ministry of Education and Science of
Russia (contract \#16.513.11.3036).

\end{document}